\documentclass[12pt,epsf,epsfig]{article}
\usepackage{amsmath}
\usepackage{graphicx}

\begin{document}

\begin{center}
{\huge{Brane world creation from flat or almost flat space in dynamical tension string theories}}  \\
\end{center}

\begin{center}
 E.I. Guendelman  \\
\end{center}

\begin{center}
\ Department of Physics, Ben-Gurion University of the Negev, Beer-Sheva, Israel \\
\end{center}

\begin{center}
\ Frankfurt Institute for Advanced Studies, Giersch Science Center, Campus Riedberg, Frankfurt am Main, Germany \\
\end{center}

\begin{center}
\ Bahamas Advanced Studies Institute and Conferences,  4A Ocean Heights, Hill View Circle, Stella Maris, Long Island, The Bahamas \\
\end{center}
(corresponding author, e-mail:  guendel@bgu.ac.il),     
\begin{center}
 J. Portnoy  \\
\end{center}
\begin{center}
\ Bahamas Advanced Studies Institute and Conferences,  4A Ocean Heights, Hill View Circle, Stella Maris, Long Island, The Bahamas \\
\end{center}

\abstract
There is great interest in the construction of brane worlds, where matter and gravity are forced to be effective only in a lower dimensional surface , the ¨brane¨ . How these could appear as a consequence of string theory  is a crucial question and this has been widely discussed. Here we will examine a distinct scenario that appears in dynamical tension theories and where string tension is positive between two surfaces separated by a short distance and  at the two surfaces themselves the string tensions become infinite, therefore producing an effective confinement of the strings and therefore of all matter and gravity to the space between these to surfaces, which is in fact a new type of stringy brane world scenario. The basic formulation for obtaining this scenario consist of assuming two types of strings characterized by a different constant of integration related to the spontaneous string tension generation. These string tension multiplied by the embedding metric define conformally related metrics that both satisfy Einstein´s equation.
The braneworlds appear very naturally when these two metrics are both flat spaces related by a special conformal transformation. The two types of string tensions are determined and they blow up at two close expanding surfaces. A puzzling aspect appears then: the construction is based on flat spaces, but then there are also strings with very large tension near the boundaries of the braneworld,so can the back reaction from the infinite tension strings destroy the flat space background?. Fortunatelly that can be resolved using the mechanism Universe creation from almost flat (or empty) spaces, which incorporates a gas of very large string tensions in a membrane, studied before in 1+1 membranes in a 2+1 embedding space and now is generalized for a 1+(D-2) membrane moving in a  1+(D-1) space.

\section{Introduction}

In a previous publication \cite{braneworldswithDT}  we have shown, in the context of 
theories where the string tension becomes a dynamical variable, using the modified measures formalism, which was  previously used for a certain class of modified gravity theories under the names of Two Measures Theories or Non Riemannian Measures Theories, see for example \cite{d}, \cite{b}, \cite{Hehl}, \cite{GKatz}, \cite{DE}, \cite{MODDM}, \cite{Cordero}, \cite{Hidden}
Leads to the  modified measure approach to string theory \cite{stringtheory}, where  rather than to put the string tension by hand it appears dynamically.

This approach has been studied in various previous works  \cite{a,c,supermod, cnish, T1, T2, T3}. See also the treatment by Townsend and collaborators for dynamical string tension \cite{xx,xxx}. 

In  \cite{braneworldswithDT}, and in \cite{ESSAY} and references there
we have also introduced the ¨tension scalar¨, which is an additional
background
field that can be introduced into the theory for the bosonic case (and expected to be well defined for all types of superstrings as well) that changes the value of the tension of the extended object along its world sheet. Before studying issues that are very special of this paper we review some of the material contained in previous papers,  first present the string theory with a modified measure and containing also gauge fields that like in the world sheet, the integration of the equation of motion of these gauge fields gives rise to a dynamically generated string tension, this string tension may differ from one string to the other.

Then we consider the coupling of gauge fields in the string world sheet to currents in this world sheet, as a consequence this coupling induces variations of the tension along the world sheet of the string. Then we consider a bulk scalar and how this scalar naturally can induce this world sheet current that couples to the internal gauge fields. The integration of the equation of motion of the internal gauge field lead to the remarkably simple equation that the local value of the tension along the string is given by $T= e \phi + T _{i} $ , where $e$ is a coupling constant that defines the coupling of the bulk scalar to the world sheet gauge fields and  $ T _{i} $ is an integration constant which can be different for each string in the universe. 

Then each string is considered as an independent system that can be quantized. We take into account the string generation by introducing the tension as a function of the scalar field as a factor inside a Polyakov type action with such string tension, then the metric and the factor $g \phi + T _{i} $  enter together in this effective action, so if there was just one string the factor could be incorporated into the metric and the condition of world sheet conformal invariance will not say very much about the scalar  $\phi $ , but if many strings are probing the same regions of space time, then considering a background metric $g_{\mu \nu}$ , for each string the ¨string dependent metric¨  $(\phi + T _{i})g_{\mu \nu}$ appears and in the absence of othe background fields, like dilaton and antisymmetric tensor fields, Einstein´s equations apply for each of the metrics $(\phi + T _{i})g_{\mu \nu}$, considering two types of strings with $T _{1 \neq }T _{2}$. We call $g_{\mu \nu}$ the universal metric. In \cite{braneworldswithDT} the metrics $(\phi + T _{i})g_{\mu \nu}$, for $i=1, 2$ are taken to be Minkowski space and  Minkowski space after a special conformal transformation. There are then solutions  for the tensions of the two types of strings that imply a brane type, where the string tension becomes infinite at two expanding surfaces, so that all matter and gravity are constrained to be between those surfaces.

Here we want to discuss how, now from the point of view of of a gravitational theory,  this phenomenon of arbitrarily large tensions can be consistent with the existence of flat spaces.
  
   \section{ Are the Flat Space Backgrounds Consistent with the presence of very high Tension Strings?}

The whole construction of the braneworld has been based on the conformal mapping between two flat spaces, this conformal mapping then defines the behavior of the string tensions and in principle it represents a vacuum solution where test strings acquire string tensions that diverge at two concentric and expanding surfaces, for details see \cite{braneworldswithDT}.

Furthermore, as we start to populate the braneworld with actual strings, these strings will have infinite tension at the borders of the braneworld. A natural question one may ask at this point is the following :  Are the flat space backgrounds of our construction consistent with the presence of very high Tension Strings or will the backreaction from the very large string tension destroy this basic feature of the model ?. 

This question requires a non trivial answer because the presence of arbitrarily large string tensions would appear at first sight  substantial back reaction from the space time and possibly large deviations from the construction based on the flat spaces in the previous sections, but is that so? . As we will see,indeed, our picture it appears that the introduction of large tension strings is consistent with the matching of two flat or almost flat space times.
   \subsection{A General Relativistic ¨Macroscopic¨ String Gas Shells Model With Arbitrarily Large Tensions }
A most important observation in this respect is that indeed two spaces that are almost flat can be matched with a surface with matter 
described by a string gas with arbitrarily large tensions \cite{universesfromflatspace}. There are obstacles to directly compare the braneworld solutions in dynamical tension strings and those found in \cite{universesfromflatspace}, which are: 1) \cite{universesfromflatspace} describes a $2+1$ dimensional brane moving in an embedding bulk space of $3+1$ dimensions, while for string theories we must consider higher dimensions, 2) in \cite{universesfromflatspace} Einstein gravity is assumed with one metric to hold in the embedding bulk space, while the effective gravity theory for the dynamical tension strings two string metrics appear, 3) in \cite{universesfromflatspace} an infinitely thin brane is considered, while in  dynamical tension strings the branes are thick, that is why the thin wall model will be referred as a ¨macroscopic¨ representation of the braneworld scenario.

A difference can be resolved in a simple way is generalizing the dimensions of the brane to  $(D-2)+1$ and that of the embedding space to $(D-1)+1$. This we will do, while we hope the other aspects will not change the basic qualitative aspects of the comparison.

We consider then a surface or thin shell with  $D-2$ spacial dimensions, where in this shell a gas of strings with the equation of state that relates the surface pressure $p$ to the  $\sigma$ being
\begin{equation}\label{string gas}
p= - \frac{\sigma}{D-2}
\end{equation}
see for example a discussion of the string gas equation of state in $4D$ cosmology in \cite{stringgasequationofstate})
and for an example involving string gas shells see \cite{StringGasShells},
so for $D=3$, we obtain that the surface becomes a line with $p= - \sigma $, This was a matching corresponding to a particular choice of the ones studied in 
\cite{Jacob1}, while the  $D=4$ corresponds to a membrane ($2+1$ dimensional brane) moving in $3+1$ universe with a string gas matter in it \cite{universesfromflatspace}. In  \cite{stringgasequationofstate} and in \cite{Jacob1} the universe was meant to be the bulk space inside the bubble, while now, being interested in the braneword picture, the bubble, that is the surface with the large string tensions itself is the Universe where we live. We must consider therefore higher dimensions to get a relevant braneworld scenario. 

Applying a local conservation law of the energy momentum in the brane defined by eq. (\ref{string gas}) leads to the possibility of integrating $\sigma$,
\begin{equation}\label{sigma}
\sigma= \frac{\sigma_0}{r^{D-3}}
\end{equation}
where $\sigma_0$ is a constant. As we can see, for $D=3$, $\sigma=$ constant, as  considered as a particular case in \cite{Jacob1} while for $D=3$, $\sigma= \frac{\sigma_0}{r}$ as  considered  in \cite{universesfromflatspace}.
These cases used the Israel matching conditions \cite{Israel} for two space times separated by the string gas shell that we will now generalize to higher dimensions. Following \cite{universesfromflatspace} generalized now to higher dimensions, we consider two stationary metrics with rotational invariance of the form,
\begin{equation}\label{outsidemetric}
ds^2= -A_{+}dt^2 +\frac{dr^2}{A_{+}} +r^2 d\Omega^2_{D-2}
\end{equation}
for the outside metric and
\begin{equation}\label{insidemetric}
ds^2= -A_{-}dt^2 +\frac{dr^2}{A_{-}} +r^2 d\Omega^2_{D-2}
\end{equation}
for the inside metric. Here $d\Omega^2_{D-2}$ represents the contribution to the metric of the $D-2$ angles relevant to the spherically symmetric solutions in $D$ space time dimensions.
$A_{+}$ and $A_{-}$ are functions of $r$ , different for the inside and the outside spaces, matched at a bubble defined by a trajectory
\begin{equation}\label{trajectory}
r= r(\tau)
\end{equation}
Then the matching condition as a consequence of the Israel analysis \cite{Israel} generalized to $D$ dimensions reads,
\begin{equation}\label{matching}
\sqrt{A_{-} + \dot{r}^2}-\sqrt{A_{+} + \dot{r}^2} = \kappa \sigma r
\end{equation}
where $\kappa$ is proportional to Newton constant in $D$ dimensions.
The square roots are not necessarily positive, the sign can be negative for example for a wormhole matching as has been discussed in details in $D=4$, which corresponds to a membrane ($2+1$ dimensional brane) moving in $3+1$ universe with a string gas matter in it \cite{universesfromflatspace}. Another case where a difference a sum of the  two terms is obtained, or what is equivalent, we can say that  the second square root is negative is when considering a braneword scenario where the radius growths as we go out from the brane on both sides, see for example  \cite{DynamicsofAnti-deSitterDomainWalls} .
The assignment of signs of the square roots when one of the soaces is a Schwarzschild space can be worked out rigorously by study the problem using Kruskal–Szekeres  coordinates \cite{BlauGuendelmanGuth}
where these expressions were used for the study of the dynamics of false vacuum bubbles and baby universe creation

We will now study the case where inside we have flat space, that is

 $A_{-} =1$

and outside a $D$ dimensional Schwarzschild solution with maximal rotational invariance, which gives the Tangherlini solution \cite{Tangherlini}

$A_{+} = 1- \frac{c_{1}}{r^{D-3}} $

where $c_{1}$ is a constant. In the Tangherlini solution the radial fall off $\frac{1}{r}$ of the Newtonian potential is replaced by the $\frac{1}{r^{D-3}}$ behavior. These expressions have been used for $D=3$ in \cite{Jacob1}, while the  $D=4$ was studied in  \cite{universesfromflatspace}, we now generalize for any dimension.

Solving from \ref{matching} for one of the square roots and then solving for the other square root and squaring again, we obtain the particle in a potential like equation,
\begin{equation}\label{particlelikeequation}
\dot{r}^2 + V_{eff}(r)= 0
\end{equation}

where 

\begin{equation}\label{effectivepotential}
 V_{eff}(r)= - (\frac{r^{D-4}}{\kappa \sigma_0}-
 \frac{c_1}{2\kappa \sigma_0 r} - \frac{\kappa \sigma_0}{2} r^{-D+4} )^2 +
 \frac{r^{2D-8}}{(\kappa \sigma_0)^2} -  \frac{c_1 r^{D-5}}{(\kappa \sigma_0)^2}
\end{equation}

It is  useful to look at the solutions where $ V_{eff}(r)= 0$,
 from (\ref{effectivepotential}) and considering $c_1=0$, so we are matching two flat spaces, we get,
 \begin{equation}\label{Vzero}
 V_{eff}(r)= -(\frac{r^{D-4}}{\kappa \sigma_0}
  - \frac{\kappa \sigma_0}{2} r^{-D+4} )^2 +
 \frac{r^{2D-8}}{(\kappa \sigma_0)^2} = 0
\end{equation}
 
 which is solved by
  \begin{equation}\label{Vzero2sols}
  - (\frac{r^{D-4}}{\kappa \sigma_0}-
  - \frac{\kappa \sigma_0}{2} r^{-D+4}  )^2 = -
 \frac{r^{2D-8}}{(\kappa \sigma_0)^2}
\end{equation}

multiplying by minus one and taking the minus square root in the right hand side, we get
  \begin{equation}\label{Vzero2solsonewith minus}
  (\frac{r^{D-4}}{\kappa \sigma_0}
  - \frac{\kappa \sigma_0}{2} r^{-D+4}  ) = -
 \frac{r^{D-4}}{(\kappa \sigma_0)}
\end{equation}
which leads to the solution for the maximum radius,

 \begin{equation}\label{solforraduis}
  r_m= 
 (\frac{(\kappa \sigma_0)^2}{4})^{^{\frac{1}{2D-8}}}
 = 
 (\frac{\kappa \sigma_0}{4})^{^{\frac{1}{D-4}}}
 \end{equation}
 for large $D$, that is $D$ bigger than four, the exponent is positive, so that the maximum radius goes to infinity as the string tensions goes to infinity.
 
The expression (\ref{effectivepotential}) can also be expressed as
\begin{equation}\label{effectivepotentialalternative}
 V_{eff}(r)=1 - (
 \frac{c_1}{2\kappa \sigma_0 r} +  \frac{\kappa \sigma_0}{2r^{D-4}}  )^2
\end{equation}
This expression coincides with the expression obtained in \cite{universesfromflatspace} for the case where $D=4$, as we see from the above expression, the potential, even for $c_1 \neq 0$ goes to a constant
as $r \rightarrow \infty $,
as depicted for a particular choice of parameters in FIG1,
\begin{center}\includegraphics[scale=.2]{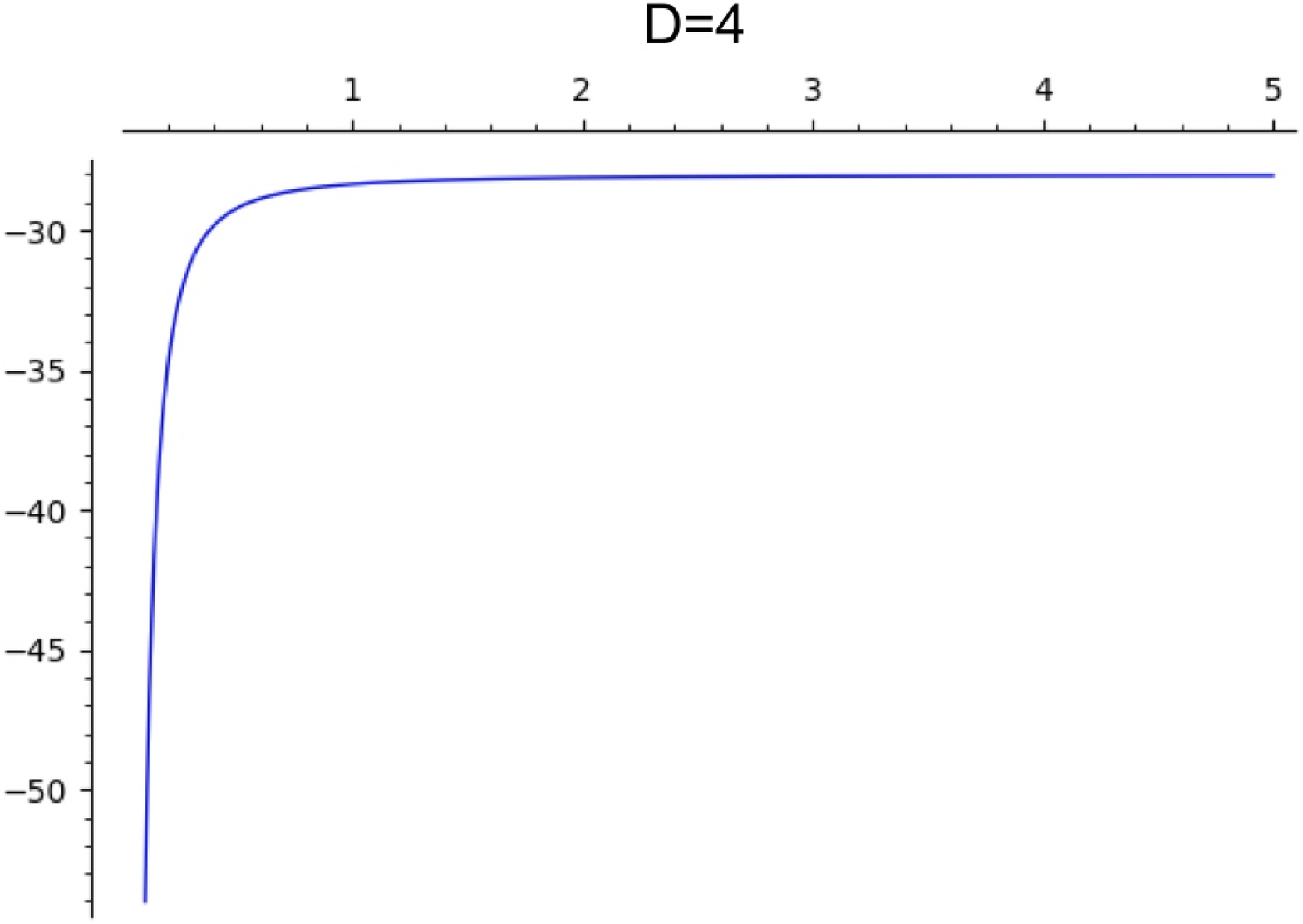}\end{center} 
\begin{center} Fig1. The potential for a particular choice of parameters in D=4\end{center}
 This constant can be positive or negative depending on whether 
$\kappa \sigma_0$ is big or small, for $\kappa \sigma_0 > 2 $  the asymptotic value of the potential is negative and the membrane approaches $\infty $ with constant velocity
regardless of the value of $c_1$, so large string tensions produce indeed child universes even in the case the inside and outside spacetimes are flat ( the inside space time is always flat, while the outside space is flat for $c_1 = 0$). The matching requires a wormhole as shown in \cite{universesfromflatspace}.
For D=3 we have that for $r \rightarrow 0$ and   $r \rightarrow  \infty$ the potential is negative (see Fig. 2), or in the case of more interest to us, if $c_1=0$ in the limit $\sigma_0 \rightarrow \infty$ it will be negative everywhere except for an infinitesimal region around r=0
 \begin{center}\includegraphics[scale=.3]{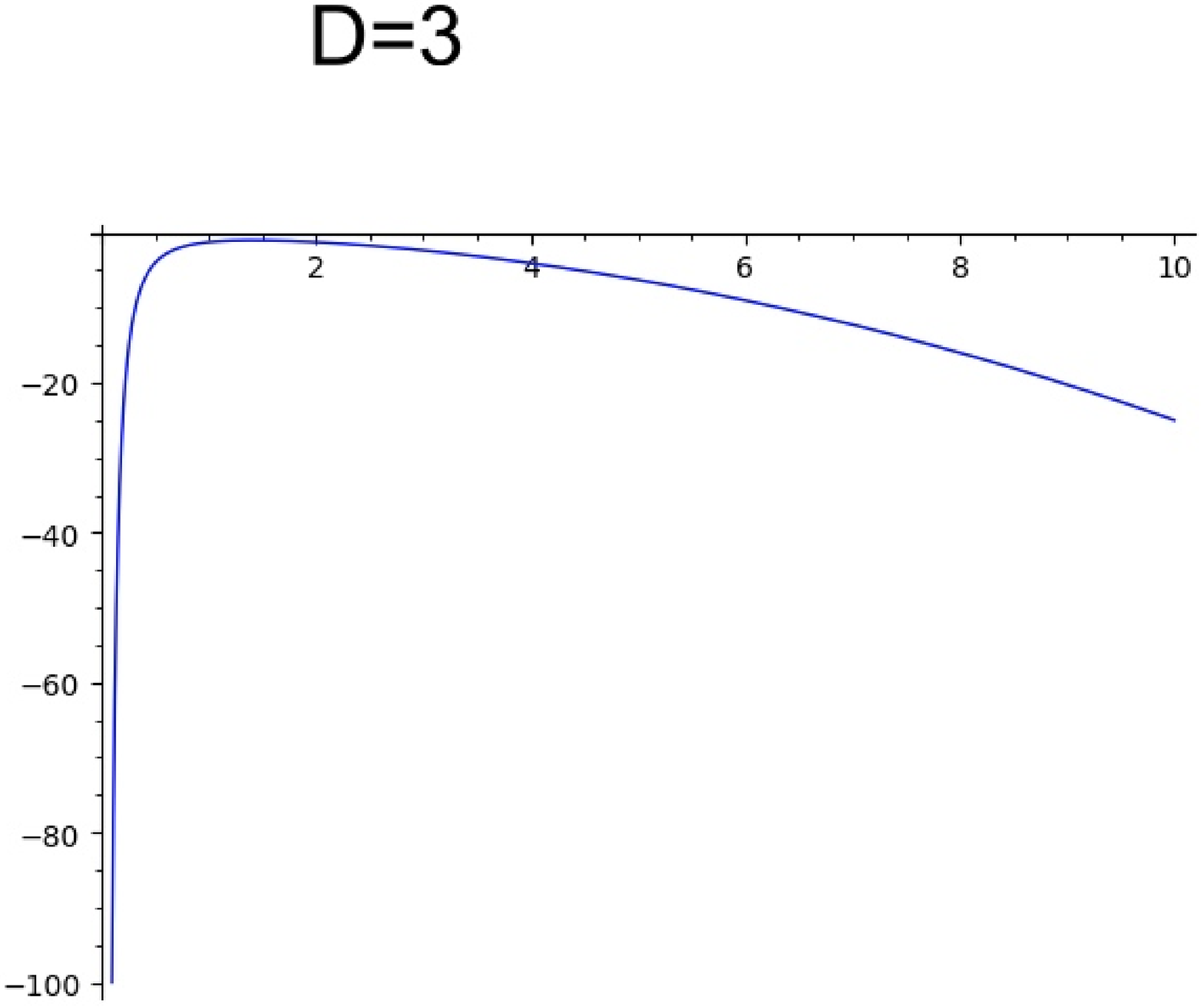}\end{center} 
 \begin{center} Fig2. The potential for a particular choice of parameters in D=3\end{center}
The expression (\ref{effectivepotentialalternative}) allows also a particularly simple solution for the point where  $ V_{eff}(r)= 0$, the point of return of the bubble,  even for $c_1 \neq 0$, which is particularly simple for $D=5$, see FIG3 since in this case both terms inside the square  are proportional to $\frac{1}{r}$, 
so we must choose

\begin{equation}\label{effectivepotentialzerofordimension5}
 1 - 
 (\frac{c_1}{2\kappa \sigma_0 r} +  \frac{\kappa \sigma_0}{2r}) =
  1-\frac{1}{r} (\frac{c_1}{2\kappa \sigma_0} + \frac{\kappa \sigma_0}{2})= 0
 \end{equation} 
 \begin{center}\includegraphics[scale=.7]{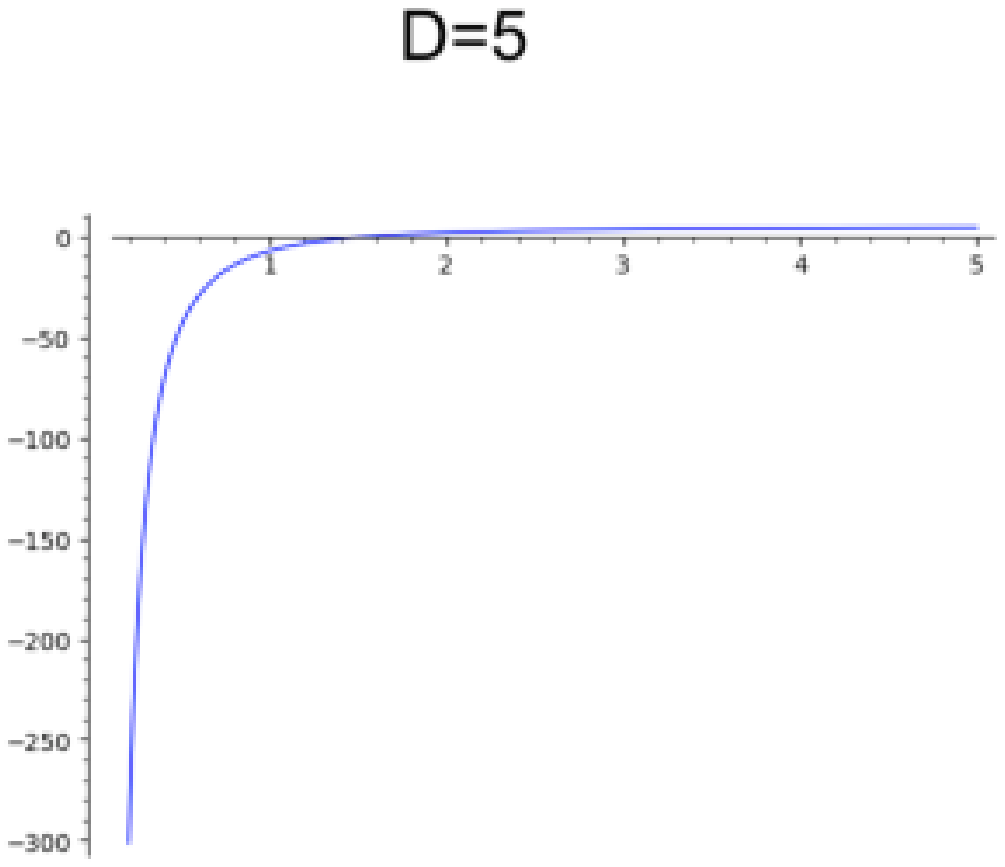}\end{center} 
 \begin{center} Fig3. The potential for a particular choice of parameters in D=3\end{center}
 so that  
 \begin{equation}\label{rmaximumfordimension5}
 r_m =
  (\frac{c_1}{2\kappa \sigma_0} + \frac{\kappa \sigma_0}{2})
 \end{equation} 
 
 So, we explicitly see that for
  $\kappa \sigma_0   \rightarrow \infty$ , then , 
 $ r_m \rightarrow \infty $ regardless of the mass (i.e. $c_1$),
 so infinite tension string gas shell can describe an expanding shell to infinity being connected by two flat spaces. This feature extends to all dimensions bigger than $4$ as well. Finally, FIG 4 shows the effective potential for $D=26$.
 \begin{center}\includegraphics[scale=.7]{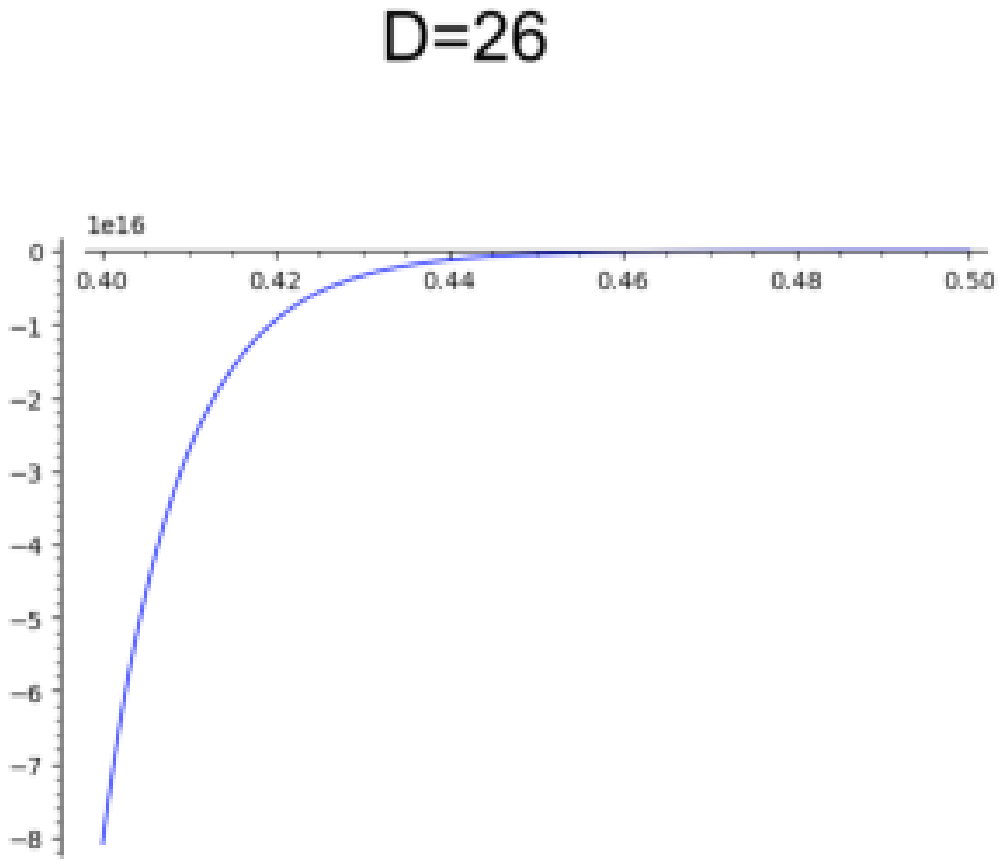}\end{center} 
 \begin{center} Fig4. The potential for a particular choice of parameters in D=26\end{center}
  
 \subsection{Further Possible Improvements on the Macroscopic General Relativistic string gas model with arbitrarily large tensions }
 The toy model described in the previous section describes some of the rough features of what is taking place in the braneworld scenario, in particular the presence of arbitrarily large string tensions without giving up the flat space structure of the background metrics. 
 
 Although this basic feature is captured by the model, it is clear that this is only a first step, since 1) the braneworld does not start as a thin wall, this happens only at asymptotically large times, 2) in the real braneworld scenario, the internal wall is initially missing,
 the universe starts as a ball where the internal wall appears and then becomes very close to the outside wall, which suggests that the tensions will in fact grow in time, this should help the expansion of the shell to infinity in the gravity picture of the scenario, which require indeed large string tensions to expand to infinity.
 These more refined models will be studied in a future paper.

\textbf{Acknowledgments}
 We thank  Stefano Ansoldi  for usefull discussions and collaborations on the subject of Universes out of almost flat spaces
 and Zeeya Merali for additional collaborations on the subject of baby universe creation. EG also want to thank the Foundational Questions Institute (FQXi)  and the COST actions  Quantum Gravity Phenomenology in the multi messenger approach, CA18108 and  Gravitational waves, Black Holes and Fundamental Physics, CA16104 for support.


\begin{thebibliography}{9}
\bibitem{braneworldswithDT}
Light like segment compactification and braneworlds with
dynamical string tension, E.I. Guendelman, Eur. Phys. J. C (2021) 81:886
https://doi.org/10.1140/epjc/s10052-021-09646-z
\bibitem{stringtheory}  ¨Superstrings¨,
John H. Schwarz, Vols 1 and 2, World Scientific, 1985; M. B. Green, J. H. Schwarz and E. Witten, Superstring Theory, Cambridge University
Press, 1987.
\bibitem{d}
E.I. Guendelman, A.B. Kaganovich, Phys.Rev.D55:5970-5980 (1997)
\bibitem{b}
E.I. Guendelman, Mod.Phys.Lett.A14, 1043-1052 (1999)
\bibitem{GKatz} E.I. Guendelman, O. Katz,  Class.Quant.Grav. 20 (2003) 1715-1728 • e-Print: gr-qc/0211095 [gr-qc]
\bibitem{Hehl}
Frank Gronwald, Uwe Muench, Alfredo Macias, Friedrich W. Hehl, Phys.Rev.D 58 (1998) 084021 • e-Print: gr-qc/9712063 [gr-qc]
\bibitem{DE}
Eduardo Guendelman, Ramón Herrera, Pedro Labrana, Emil Nissimov, Svetlana Pacheva, Gen.Rel.Grav. 47 (2015) 2, 10 • e-Print: 1408.5344 [gr-qc]
\bibitem{MODDM}
Eduardo Guendelman, Douglas Singleton, Nattapong Yongram, JCAP 11 (2012) 044 • e-Print: 1205.1056 [gr-qc]
\bibitem{Cordero}
R. Cordero, O.G. Miranda, M. Serrano-Crivelli, JCAP 07 (2019) 027 • e-Print: 1905.07352 [gr-qc]
\bibitem{Hidden}
Eduardo Guendelman, Emil Nissimov, Svetlana Pacheva, Eur.Phys.J.C 75 (2015) 10, 472 • e-Print: 1508.02008 [gr-qc]
\bibitem{a}
E.I. Guendelman, Class.Quant.Grav. 17, 3673-3680 (2000)
\bibitem{c}
E.I. Guendelman, A.B. Kaganovich, E.Nissimov, S. Pacheva, Phys.Rev.D66:046003 (2002)
\bibitem{supermod}
E.I. Guendelman, Phys.Rev.D 63 (2001) 046006 • e-Print: hep-th/0006079 [hep-th]
\bibitem{cnish}
Hitoshi Nishino, Subhash Rajpoot, Phys.Lett.B 736 (2014) 350-355
e-Print: 1411.3805 [hep-th].
\bibitem{T1}
T.O. Vulfs, E.I. Guendelman, Annals Phys. 398 (2018) 138-145 • e-Print: 1709.01326 [hep-th]
\bibitem{T2}
T.O. Vulfs, E.I. Guendelman, Int.J.Mod.Phys.A 34 (2019) 31, 1950204 • e-Print: 1802.06431 [hep-th]
\bibitem{T3}
T.O. Vulfs, Ben Gurion University Ph.D Thesis, (2021), arXiv:2103.08979. 
\bibitem{ESSAY} Implications of the Spectrum of Dynamically Generated String Tension Theories,   E.I. Guendelman, Int.J.Mod.Phys.D 30 (2021) 14, 2142028 • e-Print: 2110.09199 [hep-th], which reviews  results presented at Escaping the Hagedorn Temperature in Cosmology and Warped Spaces with Dynamical Tension Strings, E.I. Guendelman, e-Print: 2105.02279 [hep-th] and  Cosmology and Warped Space Times in Dynamical String Tension Theories, Eduardo Guendelman, e-Print: 2104.08875 [hep-th]

\bibitem{xx}
P.K. Townsend,  Phys.Lett.B 277 (1992) 285-288.
\bibitem{xxx}
E. Bergshoeff, L.A.J. London, P.K. Townsend,
Class.Quant.Grav. 9 (1992) 2545-2556, Class. Quantum Grav. 9 (1992) 2545-2556  • e-Print: hep-th/9206026 [hep-th]
\bibitem{pol1} Deser, S. and Zumino,Phys. Lett. B65 , 369, (1976)
\bibitem{pol2} Brink , L., Di Vechia, P and Howe, S. ,Phys. Lett. B65, 471 , (1976)
\bibitem{pol3} Polyakov, A. M,, ,Phys. Lett. B103 ,207, (1980).
\bibitem{Schwinger} Particles and Sources,
Julian Schwinger,  Phys.Rev. 152 (1966) 1219-1226
DOI: 10.1103/PhysRev.152.1219
\bibitem{Ansoldi}
S. Ansoldi, E. I. Guendelman, E. Spallucci, Mod.Phys.Lett.A 21 (2006) 2055-2065 • e-Print: hep-th/0510200 [hep-th]
\bibitem{cosmicstrings}
A. Vilenkin, E.P. S. Shellard, Cosmic Strings and Other Topological Defects, Cambridge Monographs on Mathematical Physics,  Cambridge University Press,  2000,  580 pages,
ISBN: 9780521654760.
\bibitem{Polchinski} Joseph Polchinski, String Theory, vol. 1 , Cambridge University Press (1998) ; some papers on strings with background fields are C. G.
Callan, D. Friedan, E. J. Martinec and M. J. Perry, Nucl. Phys. B 262 (1985) 593;
T. Banks, D. Nemeschansky and A. Sen, Nucl. Phys. B 277 (1986) 67
\bibitem{Culetu} H. Culetu,
The special conformal transformation and Einstein's equations, 
 Il Nuovo Cimento B, 621–628 (1989)
\bibitem{Kastrup}
On the Advancements of Conformal Transformations and their Associated
Symmetries in Geometry and Theoretical Physics1,  H.A. Kastrup,
Annalen Phys.17:631-690,2008, arXiv:0808.2730 [physics.hist-ph].
\bibitem{stringgasequationofstate}
Some properties of the 'String gas' with the equation of state $ p = -\rho/3 $, 
Alexander Yu. Kamenshchik, Isaak M. Khalatnikov, Int.J.Mod.Phys.D 21 (2012) 1250004 • e-Print: 1109.0201 [gr-qc]
\bibitem{StringGasShells}String Gas Shells, their Dual Radiation and Hedgehog Signature Control, 
E.I. Guendelman, Phys.Lett.B 677 (2009) 71-73 • e-Print: 0903.2127 [gr-qc].
\bibitem{universesfromflatspace}Universes out of almost empty space, Stefano Ansoldi, Eduardo I. Guendelman, Prog.Theor.Phys. 120 (2008) 985-993 • e-Print: 0706.1233 [gr-qc].
\bibitem{Jacob1} Almost Classical Creation of a Universe, E.I. Guendelman and J. Portnoy,  Mod.Phys.Lett.A 16 (2001) 1079-1087
\bibitem{DynamicsofAnti-deSitterDomainWalls} Dynamics of Anti-deSitter Domain Walls, Per Kraus, JHEP 9912 (1999) 011, DOI:	10.1088/1126-6708/1999/12/011
\bibitem{Israel} Singular hypersurfaces and thin shells in general relativity,  W. Israel,  Nuovo Cim.B 44S10 (1966) 1, Nuovo Cim.B 48 (1967) 463 (erratum), Nuovo Cim.B 44 (1966) 1.
\bibitem{BlauGuendelmanGuth} The Dynamics of False Vacuum Bubbles
S. K. Blau, E.I. Guendelman, A. H. Guth, Phys.Rev.D 35 (1987) 1747.
\bibitem{Tangherlini} Schwarzschild field in n dimensions and the dimensionality of space problem, F.R.Tangherlini, Nouvo Cim 27, 636 (1963). 


\end{thebibliography}
\end{document}